\documentclass[twocolumn,preprintnumbers,11pt,tightenlines,nofootinbib,superscriptaddress,noshowpacs,showkeys,aps, pra]{revtex4-2}

\usepackage{graphicx}
\usepackage{dcolumn}
\usepackage{bm}
\usepackage{url}
\usepackage{hyperref}

\usepackage{amsmath}
\usepackage{amssymb}
\usepackage[T2A]{fontenc}
\usepackage[utf8]{inputenc}
\usepackage[english]{babel}
\usepackage{multirow}
\usepackage{xcolor}
\usepackage{booktabs}
\usepackage{changes}
\usepackage{longtable}
\usepackage[normalem]{ulem}
\allowdisplaybreaks
\makeatletter

\renewcommand\thetable{\arabic{table}}

\makeatother

\renewcommand\thetable{\arabic{table}} 

\graphicspath{{figs/}}

\begin{document}


\title{Photometric and spectral variability of the candidate post-AGB  star IRAS~07506-0345}

\author{\firstname{N.~P.}~\surname{Ikonnikova}}
\email{ikonnikova@gmail.com} \affiliation{Sternberg Astronomical Institute, Lomonosov Moscow State University, Universitetski pr. 13, Moscow, 119234, Russia}

\author{\firstname{M.~A.}~\surname{Burlak}}
\affiliation{Sternberg Astronomical Institute, Lomonosov Moscow State University, Universitetski pr. 13, Moscow, 119234, Russia}

\author{\firstname{A.~V.}~\surname{Dodin}}
\affiliation{Sternberg Astronomical Institute, Lomonosov Moscow State University, Universitetski pr. 13, Moscow, 119234, Russia}

\begin{abstract}

We report photometric and spectroscopic observations of the poorly studied post-AGB candidate IRAS~07506-0345, obtained with the telescopes of the Caucasian Mountain Observatory of SAI MSU in 2023--2026. The star exhibits Cepheid-like variability with a pulsation period of $P = 33.68$~days. Low-resolution spectra reveal a strong CN molecular band at $\lambda 3883$, characteristic of carbon-rich post-AGB objects, while H$\alpha$ shows phase-dependent emission and P~Cyg-type profiles, indicating shock waves and mass outflow. Radial velocities are measured from multiple spectral lines. Using period--luminosity--color relations for Type II Cepheids, we estimate a luminosity of $L = 1050\pm250\,L_\odot$. The combined properties -- period, amplitude, light-curve morphology, infrared excess, low metallicity, and hydrogen line emissions -- securely classify IRAS~07506-0345 as an RV~Tau variable.

\end{abstract}

\maketitle

\section{Introduction}

Low- and intermediate-mass stars ($0.8\,M_\odot \lesssim M_{\rm ZAMS} \lesssim 8\,M_\odot$) at late evolutionary stages, after having left the asymptotic giant branch (hereafter, post-AGB) and evolving toward the planetary nebula phase, exhibit diverse observational signatures, including photometric variability and infrared (IR) excess indicative of circumstellar envelopes. Among these, pulsating RV~Tau variables are of particular interest: members of the old Galactic population with pulsation periods ranging from 20 to over 50 days (Wallerstein 2002), characterized by alternating deep and shallow minima in their light curves, low metallicity, and prominent emission lines in their spectra. These objects are in the final phase of stellar evolution, having already lost most of their mass and formed dusty structures that may appear as disks or extended envelopes. Studying such stars provides key insights into mass-loss mechanisms and chemical evolution during the terminal stages of low- and intermediate-mass stellar life.

The infrared source IRAS~07506-0345 ($\alpha = 07^{\mathrm{h}}53^{\mathrm{m}}06.9^{\mathrm{s}}, \delta = -03^{\circ} 53' 29.0''$, J2000.0) attracted attention in early studies due to its spectral energy distribution (SED) characteristic of late stages of stellar evolution. The first detailed investigation was carried out by Garc\'\i a-Lario et al. (1990), who included it in a sample of objects whose far-IR colors (25--100~$\mu$m) resemble those of known planetary nebulae. Based on near-IR ($JHK$) photometry and SED analysis, the authors concluded that the object is in a transitional phase between the AGB and post-AGB stages, preceding planetary nebula formation.

In  Blommaert et al. (1993), IRAS~07506-0345 was considered a candidate OH/IR source located beyond the solar circle. However, near-IR ($JHKLM$) photometry combined with the non-detection of OH 1612~MHz emission allowed the authors to rule out its classification as a classical AGB star, thereby confirming its association with a more advanced evolutionary stage.

Su\'{a}rez et al. (2006) obtained the first optical spectrum of the object, classified it as B9e, and included it in a sample of young stellar objects (Young). However, the star was later included in the Toru{\'n} evolutionary catalog of Galactic post-AGB and related objects (Szczerba et al. 2007) as a probable post-AGB candidate.

Photometric variability of the star was discovered through the ASAS-SN (Shappee et al. 2014; Kochanek et al. 2017) and ZTF (Chen et al. 2020) surveys. In the former, the object was assigned the variability class YSO; in the latter, CEPII.

Published data on this object remain extremely sparse. The goal of our work is to investigate its photometric and spectral properties based on original observations, determine fundamental stellar parameters, and refine its variability classification.

\section{Observations}

\subsection{\texorpdfstring{$BVR_CI_C$}{BVRCIC}-photometry}

Photometric observations were carried out with the 60-cm Ritchey--Chr\'{e}tien telescope (RC600) at the Caucasian Mountain Observatory (CMO) of the Sternberg Astronomical Institute, Lomonosov Moscow State University (SAI MSU). An Andor iKon-L CCD camera (2048~$\times$~2048 pixel format, 13.5~$\mu$m pixel size) was used as the detector. With a scale of $0.67''$/pixel, the field of view is $22' \times 22'$. The instrument is equipped with a standard set of $UBVR_{C}I_{C}$ photometric filters. A detailed description of the instrument and observing procedure is given in Berdnikov et al. (2020). Monitoring of the star was conducted in remote mode over four observing seasons (2023--2026). On each clear night, 2--3 exposures were obtained in the $BVR_{C}I_{C}$ bands. No $U$-band observations were performed, as the object proved too faint to achieve photometry with acceptable accuracy in this filter.

Suitable comparison stars were selected in the target field. They are identified in Gaia~DR3 as ID~3080581480793024384 ($\alpha = 07^{\mathrm{h}}53^{\mathrm{m}}03.16^{\mathrm{s}}$, $\delta = -03^{\circ}54'18.93''$, J2000.0) and ID~3080578560215259520 ($\alpha = 07^{\mathrm{h}}53^{\mathrm{m}}11.93^{\mathrm{s}}$, $\delta = -03^{\circ}54'56.46''$, J2000.0). Their $BV R_{C}I_{C}$ magnitudes were obtained by calibrating against SA99 standard stars (Galad\'\i-Enr\'\i quez et al. 2000): $B=15\,.\!\!^{\rm m}771$, $V=14\,.\!\!^{\rm m}904$, $R_C=14\,.\!\!^{\rm m}396$, $I_C=13\,.\!\!^{\rm m}989$ and $B=14\,.\!\!^{\rm m}799$, $V=14\,.\!\!^{\rm m}135$, $R_C=13\,.\!\!^{\rm m}729$, $I_C=13\,.\!\!^{\rm m}406$, respectively. Photometry for IRAS~07506-0345 is presented in Table~\ref{tab:photometry} (nightly mean epochs and magnitudes). Typical uncertainties were $\Delta B=0\,.\!\!^{\rm m}06$, $\Delta V=0\,.\!\!^{\rm m}023$, $\Delta R_{C}=0\,.\!\!^{\rm m}022$, $\Delta I_{C}=0\,.\!\!^{\rm m}007$.

\subsection{Spectroscopy}

Spectroscopic observations were carried out with the 2.5-m telescope at CMO of SAI MSU, using the low-resolution transient double-beam spectrograph (TDS) equipped with holographic gratings (Potanin et al. 2020). Andor Newton 940P cameras with E2V CCD42-10 detectors (512~$\times$~2048 pixel format) were used in both channels. Observations were performed with a long slit of width $1.\!\!^{\prime\prime}0$ or $1.\!\!^{\prime\prime}5$. The covered spectral range is 3500--7500~\AA. The achieved spectral resolution is 1300 in the blue channel (3500--5720~\AA) and 2500 in the red channel (5720--7500~\AA). The observation log, including dates, mid-exposure Julian dates, exposure times, signal-to-noise ratios (SNR), slit widths, and standard stars, used for flux calibration, is presented in Table~\ref{tab:sp-log}. Data reduction and analysis were performed using custom Python software described in detail in Potanin et al. (2020).

\begin{table}[h!]
\centering
\caption{Log of spectroscopic observations.}
\label{tab:sp-log}
\begin{tabular}{cccccc}
\hline
Date & JD-2400000 & $t_{exp}$& SNR & Slit & Std \\
&&s&&"&\\
\hline
2023-01-16 & 59960.87 & 600 & 42 & 1 & HIP114745 \\
2024-01-10 & 60320.46 & 600 & 107 & 1 & HD47272 \\
2024-02-07 & 60348.42 & 1200 & 96 & 1 & HIP29579 \\
2024-02-27 & 60368.34 & 1200 & 105 & 1 & HIP39556 \\
2024-03-31 & 60401.26 & 1200 & 119 & 1 & HD67593  \\
2025-02-10 & 60716.82 & 1200 & 123 & 1.5 & HIP39621 \\
\hline
\end{tabular}
\end{table}

\section{Data Analysis}

\subsection{Photometry}

The star was identified as variable in the ASAS-SN (Shappee et al. 2014; Kochanek et al. 2017) and ZTF (Chen et al. 2020) surveys. In the ASAS-SN Variable Stars Database, the object is designated ASASSN-V~J075307.40-035332.1. The catalog lists a period of $P = 33.366$~d, a mean $V$-band magnitude of $V = 15\,.\!\!^{\rm m}1$, and a variability amplitude of $\Delta V = 0\,.\!\!^{\rm m}94$; the variability type was classified as YSO (Young Stellar Object).

In the ZTF Catalog of Periodic Variable Stars (Chen et al. 2020), the object ZTFJ075307.40-035332.1 is listed with the following parameters: period $P = 33.297$~d, reference epoch $T_{0} = 2458379.923$, mean magnitudes $g = 15\,.\!\!^{\rm m}672$ and $r = 15\,.\!\!^{\rm m}038$, and variability amplitudes $\Delta g = 1\,.\!\!^{\rm m}227$ and $\Delta r = 0\,.\!\!^{\rm m}912$. The catalog contains 41 measurements in the $g$ band and 32 in the $r$ band for this object. The variability type is designated as CepII (Type II Cepheid).

\begin{figure*}
\centering
\includegraphics[width=\hsize]{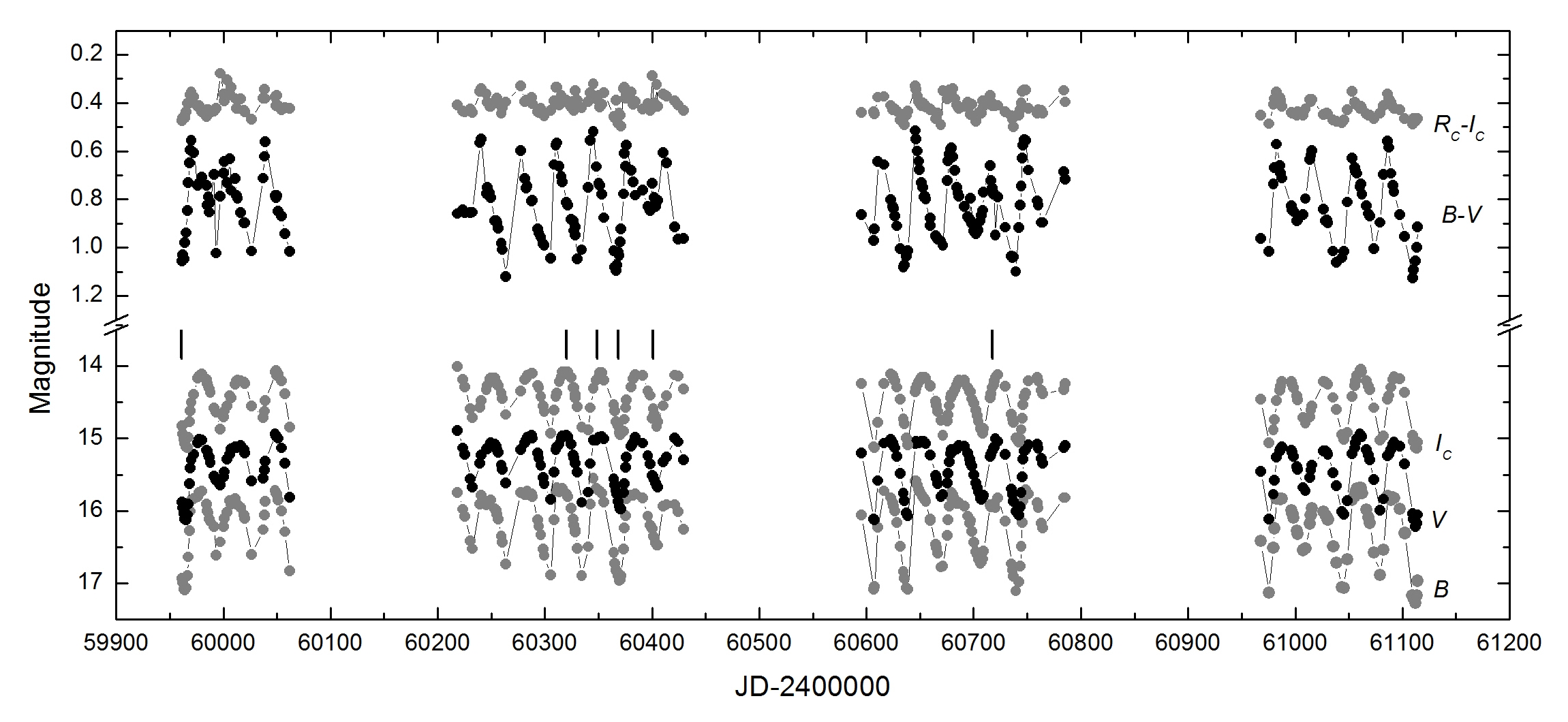}
\caption{Light curves and color indices for 2023--2026. Vertical ticks mark the epochs of spectroscopic observations.}
\label{fig:LC}
\end{figure*}

We monitored the star over four observing seasons in four photometric bands. Figure~\ref{fig:LC} presents the light curves in the $B$, $V$, and $I_{C}$ bands, along with the evolution of the $B-V$ and $R_{C}-I_{C}$ color indices during 2023--2026. Periodic variations are clearly visible in all curves. The depth of minima varies from cycle to cycle, but no strict alternation between deep and shallow minima is observed. The $B$-band light curve exhibits a somewhat asymmetric shape compared to the more symmetric profiles in the other bands: the rise to maximum is faster than the decline. The color indices also show periodic variations, though the amplitude of $R_{C}-I_{C}$ variations is substantially smaller than that of $B-V$.

Key characteristics derived from the photometric analysis -- namely, mean and maximum brightness, and peak-to-peak amplitude in the $B$, $V$, $R_{C}$, $I_{C}$ bands -- are summarized in Table~\ref{tab:photom_data}.

\begin{table}[htbp]
\centering
\caption{Photometry of IRAS~07506-0345 (2023–2026).}
\label{tab:photom_data}
\begin{tabular}{cccc}
\toprule
Band& $\langle m \rangle$& $m$(max)& $m_{\text{amp}}$\\
& mag&   mag & mag \\
\midrule
$B$      & 16.18                 & 15.55     & 1.72  \\
$V$      & 15.36                 & 14.87     & 1.34  \\
$R_C$      & 14.87                 & 14.40     & 1.21  \\
$I_C$      & 14.46                 & 13.97     & 1.17  \\
\bottomrule
\end{tabular}
\end{table}

\begin{figure}
\centering
\includegraphics[width=\hsize]{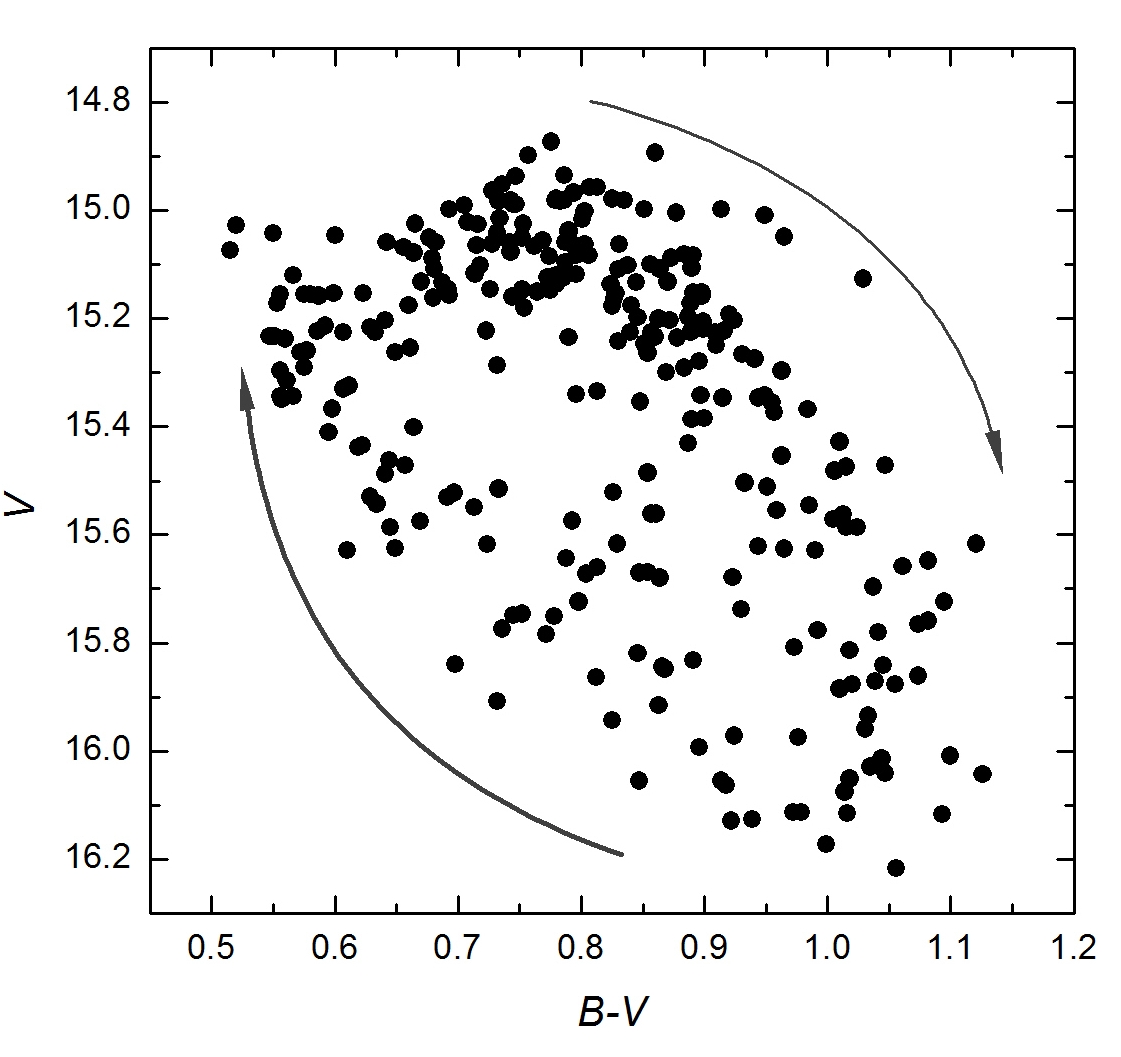}
\caption{Color--magnitude diagram. Arrows show the evolutionary direction during pulsation cycles.}
\label{fig:diagr}
\end{figure}

Figure~\ref{fig:diagr} presents the color--magnitude diagram. The star reddens as it dims. Moreover, at the same magnitude, the color is bluer on the rising branch than on the declining branch, indicating a hysteresis loop.

\subsubsection{Period analysis}

\begin{figure}
\centering
\includegraphics[width=\hsize]{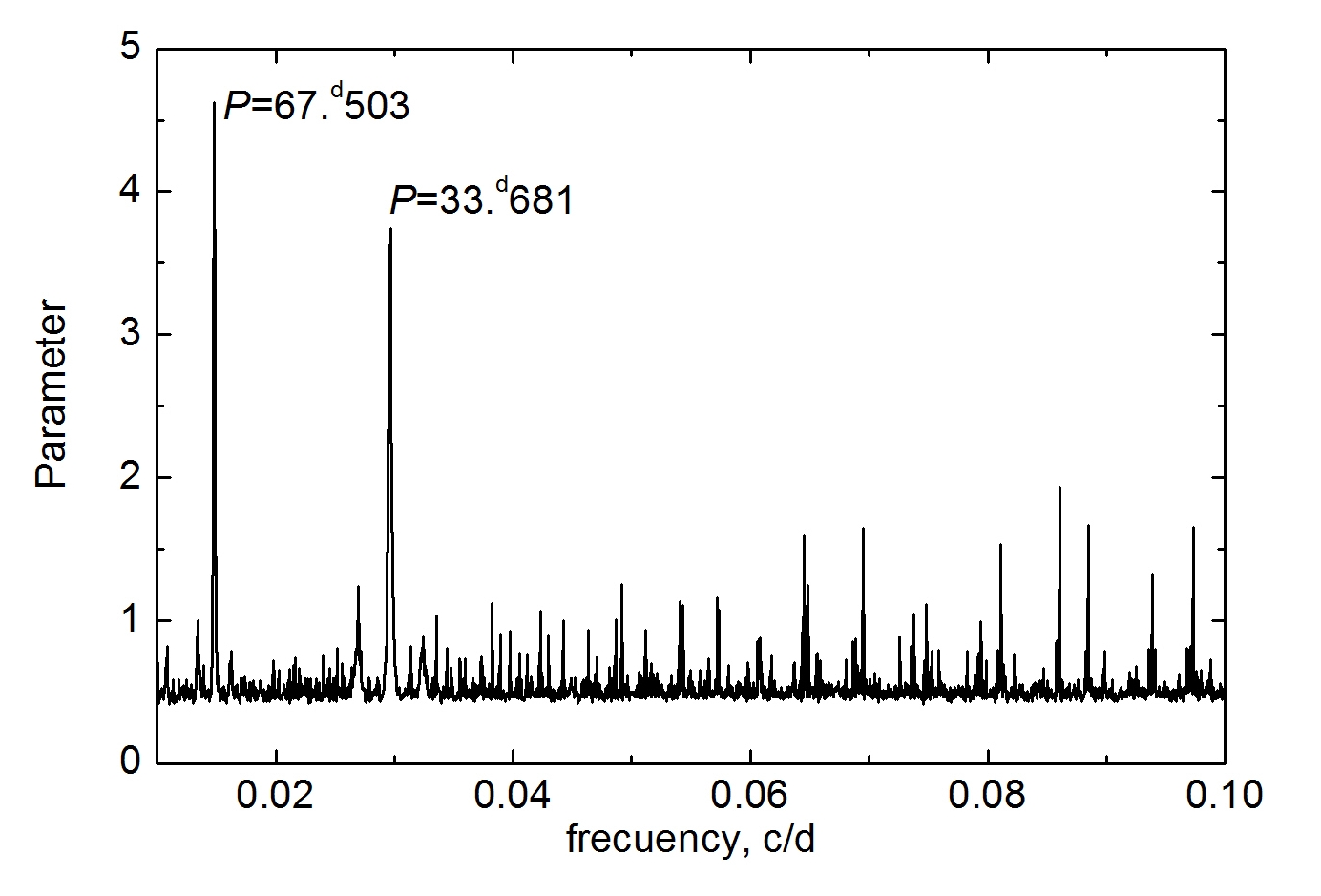}
\caption{Frequency spectrum for the $V$-band data obtained using the Lafler--Kinman method. The ordinate shows $1/\theta$, where $\theta$ is the normalized sum of squared differences between consecutive points on the phase light curve folded with a trial period. 
}
\label{fig:power}
\end{figure}

For frequency analysis, we used the WinEFK software developed by V.P.~Goranskij\footnote{The user manual is available at \url{http://www.vgoranskij.net/software/WinEFengInstruction.pdf}.}, which implements the Lafler--Kinman method (Lafler and Kinman 1965). This method minimizes the sum of squared differences between consecutive measurements when ordered by phase. The period corresponding to the minimum of the $\theta$ statistic was adopted as a preliminary estimate, followed by visual inspection of the phased light curves. Application of the Lafler--Kinman method to the $B$, $V$, $R_{C}$, $I_{C}$ data revealed peaks in the frequency spectrum corresponding to periods of $67.503$~d and $33.681$~d. Figure~\ref{fig:power} shows the frequency spectrum for our $V$-band data.

\begin{figure*}
\centering
\includegraphics[width=\hsize]{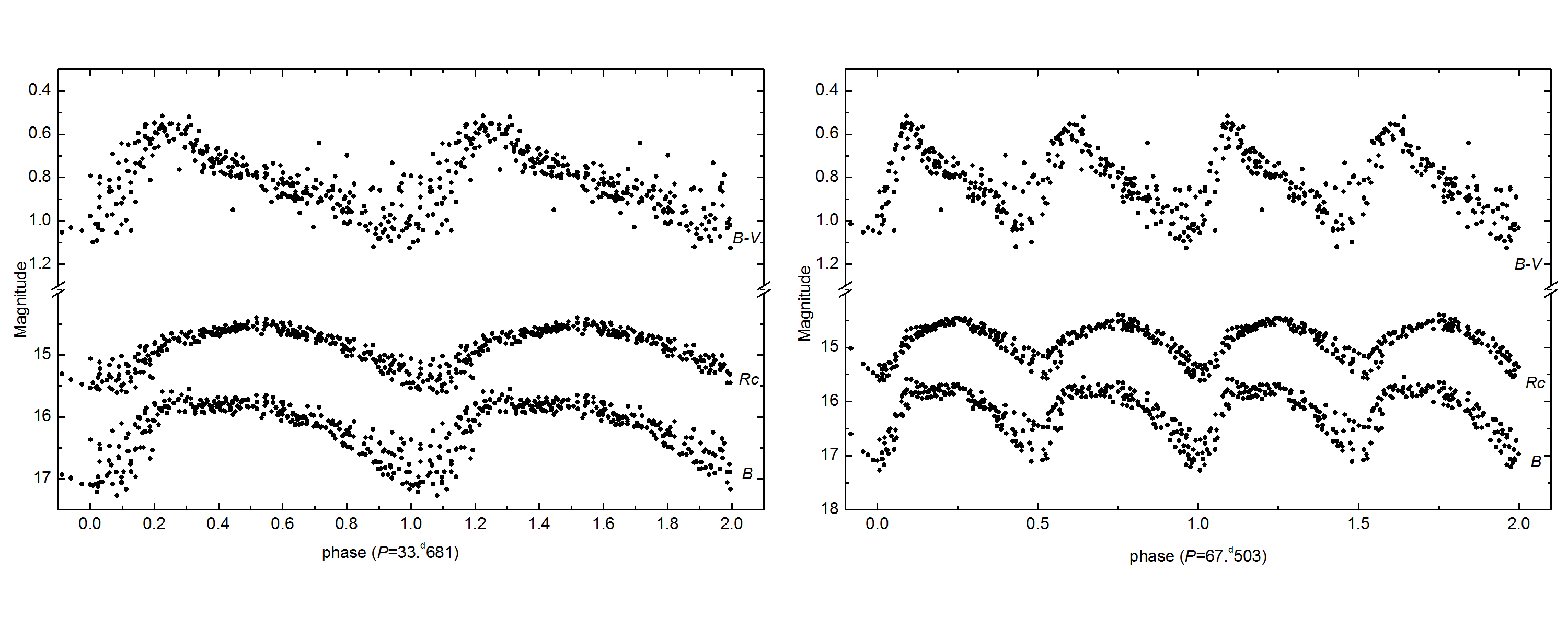}
\caption{Phased $B$, $R_{C}$ light curves and $B-V$ color curve, folded with periods of $33.681$~d and $67.503$~d.}
\label{fig:phase}
\end{figure*}

Figure~\ref{fig:phase} shows the phased light curves in the $B$ and $R_{C}$ bands, folded with periods of $33.681$~d (left panel) and $67.503$~d (right panel). The reference epoch is $T_{0} = 2459964.322$. In the phase curve folded with the doubled period, a difference between the two minima is apparent: the secondary minimum exhibits larger scatter, indicating an alternation between deep and shallow minima, though not strictly regular.

The derived periods allow us to classify IRAS~07506-0345 as an RV~Tau variable, consistent with the classification in the General Catalog of Variable Stars (Samus' et al. 2017). This class is characterized by an alternation of deep (primary) and shallow (secondary) minima in the light curve. In such cases, the formal period is defined as the interval between two consecutive deep minima, which typically exceeds the pulsation period by a factor of two, corresponding to the interval between a deep and a shallow minimum (Wallerstein 2002; Pollard et al. 1996). According Wallerstein (2002), pulsation periods of RV~Tau stars range from 20 to over 50~d.

For IRAS~07506-0345, the period derived from ASAS-SN (Shappee et al. 2014; Kochanek et al. 2017) and ZTF (Chen et al. 2020) survey data is $\approx 33.3$~d. Our observations confirm a $33.681$~d pulsation cycle, whereas the $67.503$~d interval corresponds to the formal period associated with the alternation of deep and shallow minima. The derived pulsation period ($33.681$~d) agrees well with the range typical of this class (Wallerstein 2002). 

Other photometric properties of IRAS~07506-0345 are also consistent with the characteristics of RV~Tau variables summarized in Pollard et al. (1996). Specifically, the $B$ and $V$ bands show higher amplitudes, more pronounced asymmetry, and are phase-advanced relative to the $R_{C}$ and $I_{C}$ light curves. A phase shift between the color and magnitude curves is also observed: the star reaches its bluest color not at maximum light, but on the rising branch toward maximum.

\subsection{Spectral properties}\label{sp}

To date, the only published spectral information on this star comes from Su\'{a}rez et al. (2006). The authors presented a low-resolution spectrum (wavelength range 3285--10980~\AA) obtained in March 1994 with the 1.5-m telescope at the ESO La Silla observatory. Based on these data, the star was classified as B7e and assigned to the class of young stellar objects.

Between 2023 and 2025, we obtained six spectra covering different phases of the pulsation cycle. The epochs of spectroscopic observations are marked in Figure~\ref{fig:LC}. The optical spectrum of the object is dominated by hydrogen lines of the Balmer series. The H$\alpha$ and H$\beta$ lines exhibit variable emission components.

\begin{figure}
\centering
\includegraphics[width=\hsize]{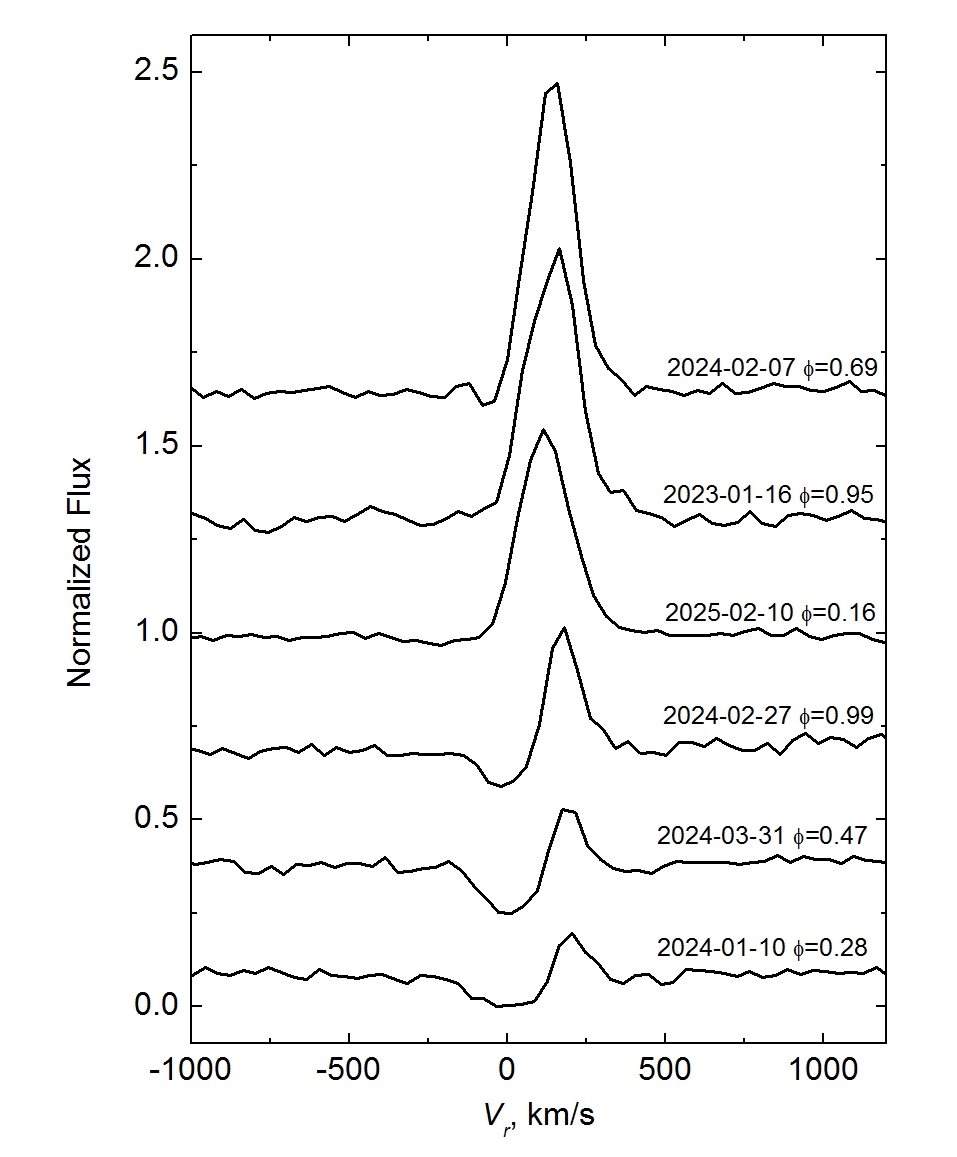}
\caption{H$\alpha$ line profiles. Observation dates and corresponding phases of the formal period ($67.503$~d) are indicated.}
\label{fig:Halpha}
\end{figure}

Figure~\ref{fig:Halpha} shows the H$\alpha$ line profiles in normalized spectra obtained at different phases of the pulsation cycle. The observed variations in the H$\alpha$ profile reflect the complex atmospheric dynamics of the star, driven by the passage of two shock waves per formal period (the interval between consecutive deep minima) (Gillet et al. 1989; Lebre and Gillet 1991, Pollard et al. 1997).

The strongest H$\alpha$ emission components are observed at phases 0.16 and 0.69. This behavior aligns with the findings of Pollard et al. (1997), who showed that RV~Tau stars typically exhibit two episodes of emission enhancement corresponding to the passage of primary (around phase 0.1--0.2) and secondary (around phase 0.6--0.7) shock waves. Notably, strong emission is also detected at phase 0.95. Based on observations of R~Sct, Lebre and Gillet (1991) demonstrated that H$\alpha$ emission driven by the secondary shock can persist up to phases $\sim$0.9--0.95. Thus, the emission observed at phase 0.95 likely represents an extended persistence of the secondary shock signature. This interpretation remains consistent with the two-shock model and reflects the object-specific characteristics of shock propagation in the atmosphere of IRAS~07506-0345.

In three spectra -- at phases 0.28, 0.47, and 0.99 -- the H$\alpha$ profile exhibits P~Cyg characteristics (red-wing emission, blue-wing absorption), providing direct spectroscopic evidence of mass outflow. Consequently, these features indicate that shock waves at these phases not only heat the atmosphere but also drive mass loss from IRAS~07506-0345.

The most prominent feature in the blue spectral region is the CN molecular band centered at $\lambda 3883$. Its intensity remains nearly constant throughout the pulsation cycle, despite variations in the stellar temperature. This stability likely indicates that the band forms in a stable, optically thick outer envelope, where temperature and density variations are insufficient to significantly alter the band intensity.

\begin{figure*}
\centering
\includegraphics[width=\hsize]{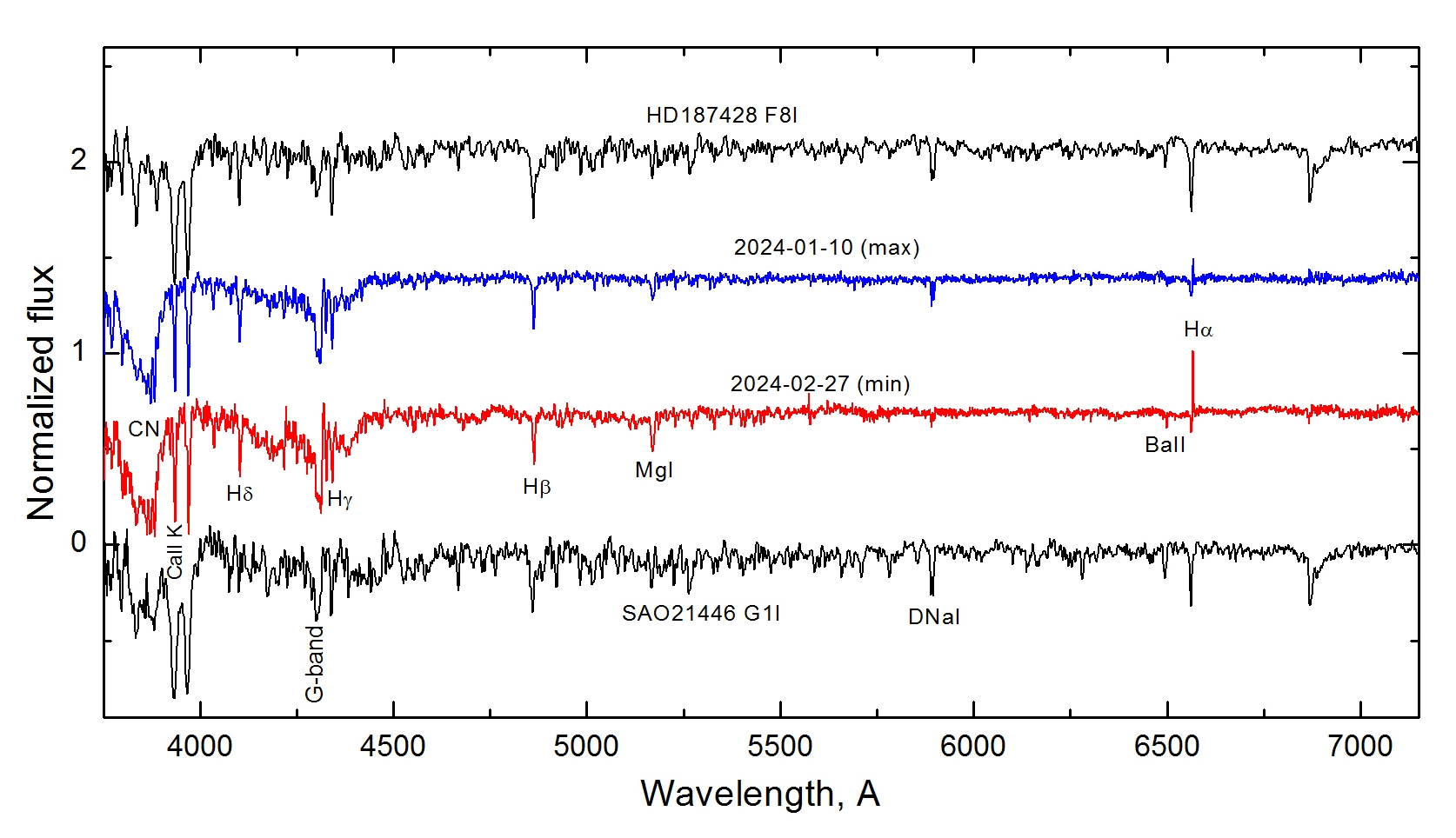}
\caption{Spectra of IRAS~07506-0345 at two pulsation phases: near minimum (red) and maximum (blue) brightness, compared with spectra of normal supergiants HD~187428 (F8I) and SAO~21446 (G1I). All spectra are continuum-normalized and arbitrarily offset along the vertical axis for clarity.}
\label{fig:sp_std}
\end{figure*}

The spectrum of IRAS~07506-0345 shows weakened metal lines compared to stars of solar metallicity. In addition to the metal-line deficiency, enhanced Ba~II lines are noticeable. No diffuse interstellar bands (DIBs) are detected in our spectra. For illustration, Figure~\ref{fig:sp_std} displays spectra of the object near minimum and maximum brightness, together with spectra of the standard stars HD~187428 (F8I) and SAO~21446 (G1I) from the library of stellar spectra (Jacoby et al. 1984).

The $B-V$ color index of IRAS~07506-0345 varies with pulsation phase from $\approx 0\,.\!\!^{\rm m}55$ at maximum brightness to $\approx 1\,.\!\!^{\rm m}1$ at minimum. Correcting for extinction with $E(B-V) = 0.03$ (see Section~4.1) yields the intrinsic color $(B-V)_{0}$ ranging from $0\,.\!\!^{\rm m}52$ to $1\,.\!\!^{\rm m}07$. Comparing these values with typical colors of normal supergiants allows us to estimate the spectral type variation over the pulsation cycle: at maximum light, $(B-V)_{0} \approx 0\,.\!\!^{\rm m}52$ corresponds to a spectral type between F5 and F8, whereas at minimum, $(B-V)_{0} \approx 1\,.\!\!^{\rm m}07$ is typical of G5 (Strai{\v{z}}ys 1992). However, the spectrum of the star exhibits a strong CN band absent in normal supergiants of comparable spectral types, along with weakened metal lines, indicating that the color-based classification should be treated with caution.

It has been repeatedly noted that the spectral classification of Type~II Cepheids using the MK system is challenging. For example, Harris and Wallerstein (1984) classified the halo-population Type~II Cepheid CC~Lyr (pulsation period $24.01$~d) as hF4mA:CH+1 at maximum light -- a notation that reflects the complex chemical composition characteristic of such objects.

Su\'{a}rez et al. (2006) classified IRAS~07506-0345 as a B7e star based on a low-resolution spectrum. However, we consider this classification incorrect. Based on our photometric analysis, the object is an RV~Tau variable. For stars of this class, the spectral type typically ranges from F to G at maximum light and from K to M at minimum (Samus' et al. 2017). Moreover, even the hottest known RV~Tau star, HP~Lyr, exhibits a spectral type of only A2--A3 at maximum light (Graczyk et al. 2002), which is significantly cooler than B7. Our photometric and spectroscopic data independently confirm a later spectral type. Thus, the B7e classification proposed by Su\'{a}rez et al. (2006) is inconsistent with both the RV~Tau variability type and our observational results, and likely stems from the low signal-to-noise ratio and insufficient spectral resolution of their observations.

\subsection{Radial velocities}

Given the low spectral resolution of our spectra and the limited number of isolated lines suitable for radial velocity measurements, we estimated this parameter and examined its possible variability.

The H$\alpha$ line exhibits a complex structure, occasionally displaying a P~Cyg profile (Figure~\ref{fig:Halpha}). Radial velocity measurements based on the emission components of H$\alpha$ reveal variations from $129.3 \pm 4.1$ to $208.9 \pm 8.9$~km\,s$^{-1}$. This observed variability is likely associated with stellar pulsations. The weighted mean from the six spectra is $151.3 \pm 2.3$~km\,s$^{-1}$.

The H$\beta$ line also shows emission components at certain epochs. Based on the two spectra exhibiting emission, the weighted mean radial velocity is $153 \pm 7$~km\,s$^{-1}$, whereas the absorption components in the remaining four epochs yield $123.6 \pm 2.8$~km\,s$^{-1}$. For the H$\gamma$ absorption line, $V_{r} = 106.6 \pm 3.0$~km\,s$^{-1}$ from the four most reliable measurements, while the weighted mean radial velocity for H$\delta$ is $128.4 \pm 5.2$~km\,s$^{-1}$.

Emission components in the lines He~I~$\lambda 5876$ and He~I~$\lambda 6678$ were detected in the spectra from 2024-02-07 and 2025-02-10, yielding radial velocities of $V_{r} = 100 \pm 3$~km\,s$^{-1}$ and $V_{r} = 114 \pm 11$~km\,s$^{-1}$, respectively.

Based on the two isolated Ba~II absorption lines at $\lambda 6141$ (five spectra) and $\lambda 6496$ (six spectra), the weighted mean radial velocities are $V_{r} = 134 \pm 4$~km\,s$^{-1}$ and $V_{r} = 134 \pm 3$~km\,s$^{-1}$, respectively.

The Na~I~D lines ($\lambda 5889.95$ and $\lambda 5895.92$) are rather weak and variable, with emission components appearing in their profiles at certain epochs. Based on absorption features measured in five spectra, we obtained weighted mean radial velocities of $V_{r} = 74 \pm 3$~km\,s$^{-1}$ and $V_{r} = 98 \pm 3$~km\,s$^{-1}$, respectively. In pulsating stars of the RV~Tau type, these lines exhibit complex profiles that may include: narrow interstellar components corresponding to the line-of-sight gas velocity and constant in time; photospheric components associated with stellar pulsations; and emission components arising in shock waves. The radial velocities derived for the two doublet lines differ significantly (74 vs.\ 98~km\,s$^{-1}$). Since lines of the same multiplet form in the same atmospheric regions, this discrepancy likely results from measurement limitations at insufficient spectral resolution: blending of interstellar, photospheric, and emission components in the complex profile shifts the measured centroid of each line.

The radial velocities derived from the Ca~II~H and K lines are $V_{r} = 144 \pm 3$~km\,s$^{-1}$ and $V_{r} = 101 \pm 3$~km\,s$^{-1}$, respectively. However, the presence of both stellar and interstellar components in these lines makes them unsuitable for deriving a precise stellar radial velocity. Furthermore, the Ca~II~H line is blended with H$\epsilon$, shifting the centroid of the blended profile redward (toward higher velocities).

Our analysis reveals that the radial velocity of IRAS~07506-0345 exhibits significant variability (from $\approx 100$ to $\approx 200$~km\,s$^{-1}$) depending on the spectral line used and the pulsation phase. The most reliable indicators of the star's photospheric velocity appear to be the Ba~II lines, which yield a stable weighted mean value of $V_{r} \approx 134$~km\,s$^{-1}$ with small uncertainty. Values derived from hydrogen and helium lines differ systematically, likely reflecting their formation in different layers of the expanding, pulsating atmosphere. The Na~I~D lines show the lowest velocities, presumably due to a substantial contribution from interstellar absorption. These results confirm the object's status as a pulsating variable of the RV~Tau type, with complex atmospheric dynamics involving expanding layers and shock waves that manifest as emission components in spectral lines.

\section{Stellar parameters}

\subsection{Luminosity, absolute magnitude, and radius}

To determine the parameters of IRAS~07506-0345, we employed empirical period--luminosity--color (PLC) relations derived for Type~II Cepheids and RV~Tau stars. Our photometric observations were carried out in the Kron--Cousins system using the $B$, $V$, $R_{C}$, $I_{C}$ filters. The relations from Alcock et al. (1998) and Manick et al. (2018), which we adopt, are calibrated in the same photometric system, as confirmed by the transformation procedures described in the original works.

While infrared-based PLC relations are generally preferred due to lower interstellar extinction in the infrared, they are unsuitable for IRAS~07506-0345. Since the object exhibits a pronounced infrared excess driven by circumstellar dust, applying such relations could introduce systematic biases into luminosity estimates. We therefore restricted our analysis to methods relying solely on optical data, which are less affected by thermal dust emission.

Our analysis is based on the PLC relation from Alcock et al. (1998), derived for 33 Type~II Cepheids and RV~Tau stars in the Large Magellanic Cloud (LMC), 
including objects with diverse dust morphologies, over the period range $0.9 < \log P < 1.75$:

\begin{equation}
\begin{split}
M_V = -0.61 (\pm 0.20) &- 2.95 (\pm 0.12) \log P  \\
&+ 5.49 (\pm 0.35) \cdot (V-R)_0,
\end{split}
\label{eq:Mv}
\end{equation}
where $P$ is the fundamental pulsation period and $(V-R)_{0}$ is the color index corrected for interstellar reddening.

As a second, independent method, we employ the PLC relation from Manick et al. (2018) for RV~Tau stars in the LMC, in the form presented by Oomen et al. (2020):

\begin{equation}
\label{eq:Mbol_Manick}
M_{\mathrm{bol}} = -3.75\cdot \log P_0 + 0.55 + BC + 2.55 (V-I)_0,
\end{equation}
where $P_{0}$ is the fundamental pulsation period and $BC$ is the bolometric correction adopted from Flower (1996). The relation was calibrated on a sample comprising both objects with circumstellar dust disks and stars without significant infrared excess, and has been successfully applied to estimate luminosities of post-AGB stars (Oomen et al. 2020).

For IRAS~07506-0345, the fundamental pulsation period is $P_{0} = 33.681$~d ($\log P_{0} = 1.527$). The mean observed color indices are $(V-R) \approx 0.50$ and $(V-I) \approx 1.07$. Using the interstellar reddening maps of Schlegel et al. (1998), with the recalibration of Schlafly et al. (2011), for the direction $l = 223.61^{\circ}$, $b = +11.84^{\circ}$, we obtain $E(B-V) = 0.027 \pm 0.001$. Applying the relations $E(V-R) = 0.79 \cdot E(B-V)$ from Alcock et al. (1998) and $E(V-I) = 1.38 \cdot E(B-V)$ from Tammann et al. (2003), we derive intrinsic colors $(V-R)_{0} \approx 0.48$ and $(V-I)_{0} \approx 1.03$.

Substituting these values into Equations~(\ref{eq:Mv}) and~(\ref{eq:Mbol_Manick}) yields:

$M_V = -2.52 \pm 0.32$, $M_{\mathrm{bol}} = M_V + BC = -2.67 \pm 0.32$, $\log(L/L_{\odot}) = 2.96 \pm 0.13$, $L = 910 \pm 280 L_{\odot}$, and

$M_{\mathrm{bol}} = -3.02 \pm 0.30$, $\log(L/L_{\odot}) = 3.10 \pm 0.12$, $L = 1260 \pm 350 L_{\odot}$, respectively.

For comparison, we also derived the luminosity using the bolometric PL relation for RV~Tau stars from Groenewegen and Jurkovic (2017) ($M_{\rm bol} = 1.442 - 2.919 \cdot \log P_{0}$, $\sigma = 0.30$), which yields $L \approx 1110~L_{\odot}$.

Three independent methods (Alcock et al. 1998; Manick et al. 2018; Groenewegen and Jurkovic 2017) yield luminosity values consistent within uncertainties, in the range $L \approx 900$--$1300~L_{\odot}$. For further analysis, we adopt the weighted mean $L = 1050 \pm 250~L_{\odot}$ ($\log(L/L_{\odot}) = 3.02 \pm 0.10$).

The stellar radius can be estimated from the Stefan--Boltzmann equation. Given that the effective temperature at maximum light is approximately 6000~K, based on the spectral classification (see Section~3.2), we obtain $R = 30 \pm 5~R_{\odot}$.

To verify the classification of IRAS~07506-0345 as an RV~Tau star, we compared our derived parameters with theoretical predictions from Bono et al. (2020). According to their evolutionary and pulsation models for Type~II Cepheids, RV~Tau stars in the post-AGB phase with masses of $0.5-0.6~M_{\odot}$ should have luminosities in the range $\log(L/L_{\odot}) = 2.5-3.2$ for pulsation periods between $20$ and $70$~d. Our derived value $\log(L/L_{\odot}) \approx 3.02$ falls within this luminosity range and corresponds to evolutionary tracks with core masses $M_{\rm c} \approx 0.53-0.55~M_{\odot}$ (Bono et al. 2020).

\subsection{Distance and halo membership}\label{dist}

In the Gaia~DR3 catalog (Gaia Collaboration 2021), the star is designated as 3080581583872232704, with coordinates $\alpha = 07^{\rm h}53^{\rm m}07.40^{\rm s}$, $\delta = -03^{\circ}53^{\prime}32.19^{\prime\prime}$ (J2000.0). The reported astrometric parameters are: parallax $\pi = 0.2021 \pm 0.0558$~mas and proper motion ${\rm PM} = 2.435$~mas~yr$^{-1}$. The distance derived from Gaia~DR3 data using Bayesian estimates (Bailer-Jones et al., 2021) is $d = 3807^{+889}_{-549}$~pc. However, we cannot consider this determination reliable. The RUWE parameter value of 2.221 indicates serious issues with the astrometric solution: for RUWE~$>1.4$, the probability that the parallax is reliable is extremely low.

Vickers et al. (2015) derived a distance of $11.18 \pm 3.24$~kpc for IRAS~07506-0345 from spectral energy distribution (SED) modeling, assuming a luminosity of $1700~L_{\odot}$. They classified the object as a halo member based on its kinematic properties and Galactic position. In this work, we obtain a different luminosity estimate, $L = 1050 \pm 250~L_{\odot}$ (see Section~\ref{results}), based on empirical PLC relations for RV~Tau stars. The discrepancy between the two luminosity values ($1050$ vs.\ $1700~L_{\odot}$) arises from different methodological approaches: our estimate relies on the individual pulsational properties of the star, whereas Vickers et al. (2015) adopted population-averaged luminosities characteristic of broad stellar groups. This underscores the importance of an individualized approach when determining parameters for RV~Tau stars.

Regardless of the exact distance, the kinematic properties of IRAS~07506-0345 indicate its membership in the Galactic halo. The measured radial velocity $V_{r} \approx +130$~km\,s$^{-1}$ significantly exceeds the expected contribution from Galactic differential rotation at latitude $b = +11.8^{\circ}$, which is characteristic of halo stars with high space velocities. This conclusion is consistent with the object's Galactic coordinates ($l = 223.6^{\circ}$, $b = +11.8^{\circ}$) and its classification as an RV~Tau star, a typical Population~II object.

\section{Discussion}

Based on the pulsation period ($33.68$~d), amplitude, and light-curve morphology, we classify IRAS~07506-0345 as an RV~Tau variable. This conclusion is further supported by the significant infrared excess, which indicates the presence of warm circumstellar dust. RV~Tau stars are typically characterized by alternating deep and shallow minima in their light curves. While IRAS~07506-0345 exhibits minima of varying depths, a strict alternation pattern is not evident. This does not contradict the classification, as this feature is known to be irregular or weakly pronounced in a subset of RV~Tau stars (Wallerstein 2002).

Based on light-curve morphology, RV~Tau stars are divided into two photometric subclasses. Class RVa comprises stars with a constant mean brightness, whereas RVb variables exhibit long-term modulation of their mean magnitude on timescales of $600-1500$~d. Neither the ASAS-SN monitoring spanning over $5000$~d (Shappee et al.; Kochanek et al. 2017) nor our observations across four observing seasons ($\Delta t > 1100$~d) reveal any variations in mean brightness. We therefore classify IRAS~07506-0345 as belonging to the RVa subclass.

The spectral features of IRAS~07506-0345, namely the presence of strong CN bands and low metallicity, are also characteristic of RV~Tau stars. However, the object cannot be unambiguously placed within the spectrophotometric classification of Preston et al. (1963). The presence of CN bands and an early spectral type (F5--F8 at maximum light) align it with Group~B, but its high radial velocity ($>100$~km\,s$^{-1}$) and membership in the Galactic halo rather than the disk contradict the criteria for this group. Conversely, the persistent presence of CN bands, which is atypical for halo objects of Group~C, rules out membership in this group as well. Thus, IRAS~07506-0345 lacks direct analogs in the Preston et al. (1963) sample, underscoring the need for further study of such objects.

The emission components of hydrogen (H$\alpha$, H$\beta$) and neutral helium lines detected in the spectrum of IRAS~07506-0345 are typical manifestations of non-stationary processes in the atmospheres of pulsating stars. Their formation is traditionally explained within the framework of the shock-wave model, whose foundations were laid by Abt (1954), Whitney (1956), and Wallerstein (1959). RV~Tau stars are characterized by the presence of two shock waves per formal pulsation cycle, which leads to two episodes of enhanced H$\alpha$ emission -- near phases $0.1-0.2$ and $0.6-0.7$ (Pollard et al. 1997).

The appearance of a P~Cyg profile in the spectrum of IRAS~07506-0345 at certain pulsation phases indicates the presence of an expanding circumstellar envelope. This is fully consistent with atmospheric dynamics models for RV~Tau stars, according to which shock waves not only produce emission features but also accelerate the outer atmospheric layers to velocities sufficient to form P~Cyg profiles (Gillet et al. 1989; Lebre and Gillet 1991).

IRAS~07506-0345 has a pulsation period of $33.68$~d, which places it among the relatively small group of RV~Tau stars with comparatively short periods. Similar objects include GK~Car ($P = 27.6$~d) and GZ~Nor ($P = 36.2$~d) (Gezer et al. 2019), as well as IRAS~02143+5852 ($P = 25$~d) (Ikonnikova et al. 2024). However, unlike these stars, IRAS~07506-0345 exhibits a less pronounced alternation of deep and shallow minima in its light curve. Nevertheless, all four objects share several key characteristics: low metallicity in their atmospheres and a significant infrared excess indicating the presence of circumstellar dust.

The WISE two-color diagram is an effective tool for classifying the nature of infrared excess. According to the criteria of Gezer et al. (2015), this diagram features a distinct region (the ``disk box'') corresponding to stars with disk-like circumstellar structures. As shown by Gezer et al. (2019), GK~Car lies within this region, whereas GZ~Nor falls outside it, indicating different morphology of its circumstellar dust environments -- for example, a multilayered envelope, as in the case of IRAS~02143+5852 (Ikonnikova et al. 2024).

The position of IRAS~07506-0345 on the WISE two-color diagram, with coordinates $[3.4]-[4.6] = 1.545$ and $[12]-[22] = 2.493$ (Cutri et al. 2013), also falls outside the ``disk box''. Thus, in terms of infrared excess characteristics, the target star shows greater similarity to GZ~Nor and IRAS~02143+5852 than to GK~Car. This suggests that the structure of its circumstellar material is not a disk but rather a multilayered envelope. 

Testing this hypothesis will require SED modeling. We plan to construct an SED model for IRAS~07506-0345 using the results of this work, new near-infrared photometric data, and archival observations from the WISE and IRAS missions. This approach will also enable an independent distance estimate based on the integrated stellar flux, without relying on population-averaged luminosity values.

An important constraint for the SED modeling is the chemical composition of the object. The presence of CN bands in the spectrum indicates that IRAS~07506-0345 is a carbon-rich (C-rich) post-AGB object. According to Verkata Raman et al. (2017), the dust envelope of such objects consists predominantly of graphite or amorphous carbon rather than silicates. Therefore, when modeling the SED of this star, one should use graphite or amorphous carbon dust, with the possible addition of silicon carbide (SiC) if a feature near $11~\mu$m is present.

The classification of IRAS~07506-0345 as an RV~Tau star naturally raises the question of its potential binarity, as numerous binary systems are known within this class. Nevertheless, we currently lack compelling evidence to consider this object a binary. First, the star exhibits no long-period photometric variations characteristic of the RVb subclass, which is frequently associated with binary systems. Furthermore, its position on the WISE two-color diagram indicates a shell-like circumstellar dust structure rather than a compact disk. In modern interpretations, the presence of a stable dust disk around post-AGB stars is generally regarded as a signature of binarity (De Ruyter et al. 2006; Oomen et al. 2018). Finally, while our spectroscopic data reveal radial velocity variability, it is most likely driven by stellar pulsations rather than orbital motion, although extended monitoring is required to draw definitive conclusions.

\bigskip

\section{Conclusions}\label{results}

We report photometric and spectroscopic observations obtained with the telescopes of the CMO SAI MSU during 2023–2026 of the poorly studied candidate post-AGB star IRAS~07506-0345. Based on our observations, combined with archival data, we present the following results:

1. Analysis of the light curves in the $B$, $V$, $R_{C}$, $I_{C}$ bands confirmed that IRAS~07506-0345 is a Cepheid-like variable star. The pulsation period has been refined to $P = 33.68$~d.

2. The combination of observational data, namely the pulsation period, light-curve morphology with varying minimum depths, the presence of a strong infrared excess, and characteristic spectral features (low metallicity, H$\alpha$ and H$\beta$ emission, and a strong CN band) enables us to confidently classify IRAS~07506-0345 as an RV~Tau star. The absence of long-period variations in mean brightness indicates its membership in the RVa subclass.

3. Using PLC relations for Type~II Cepheids, we derived a luminosity of $L = 1050 \pm 250~L_{\odot}$.

4. Although IRAS~07506-0345 exhibits features characteristic of RV~Tau stars, which are  known to contain a significant fraction of binary systems, we have found no direct evidence for binarity at this time. Arguments against binarity include the absence of long-period photometric variations (characteristic of the RVb subclass), a shell-like structure of the circumstellar dust environment, and the lack of convincing signatures of orbital radial velocity variability.

Further studies are required to conclusively determine the evolutionary status, constrain the structure of the circumstellar envelope, and test the binarity hypothesis. The primary objectives include constructing a comprehensive SED model by combining new near-infrared photometry with archival WISE and IRAS data, and continuing high-precision spectroscopic monitoring to derive a detailed radial velocity curve.

\section*{Acknowledgements}

This research was carried out within the framework of the state assignment of the Lomonosov Moscow State University.

\section*{Conflict of interest}

The authors declare no conflict of interest.

 \bigskip

REFERENCES
 \bigskip
\begin{enumerate}

\item H.A. Abt, Astrophys. J. Suppl. Ser. {\bf 1}, 63 (1954).

\item C. Alcock, R.A. Allsman, D.R. Alves, T.S. Axelrod, A. Becker, D.P. Bennett, K.H. Cook, K.C. Freeman et al., Astron. J. {\bf 115}, 1921 (1998).

\item C.A.L. Bailer-Jones, J. Rybizki, M. Fouesneau, M. Demleitner, R. Andrae, Astron. J. {\bf 161}, id. 147 (2021).

\item L.N. Berdnikov, A.A. Belinskii, N.I. Shatskii,
M.A. Burlak, N.P. Ikonnikova, E.O. Mishin,
D.V. Cheryasov, and S.V. Zhuiko, Astron. Rep. {\bf 64},
310 (2020).

\item J.A.D.L. Blommaert, W.E.C.J. Van Der Veen, and H.J. Habing, Astron. Astrophys. {\bf 267}, 39 (1993).

\item G. Bono, V.F. Braga, G. Fiorentino, M. Salaris, A. Pietrinferni, M. Castellani, M. Di Criscienzo, M. Fabrizio et al., Astron. Astrophys. {\bf 644}, A96 (2020).

\item X. Chen, S. Wang, L. Deng, R. de Grijs, M. Yang, and H. Tian, Astrophys. J. Suppl. Ser. {\bf 249}, 18 (2020).

\item R.M. Cutri, E.L. Wright, T. Conrow, J.W. Fowler, P.R.M. Eisenhardt, C. Grillmair, J.D. Kirkpatrick, F. Masci et al., VizieR Online Data Catalog, II/328 (2013).

\item S. De Ruyter, H. Van Winckel, T. Maas, T. Lloyd Evans, L.B.F.M. Waters, H. Dejonghe, Astron. Astrophys. {\bf 448}, 641 (2006).

\item P.J. Flower, Astrophys. J. {\bf 469}, 355 (1996).

\item Gaia Collaboration: A.G.A. Brown, A. Vallenari, T. Prusti, J.H.J. de Bruijne, C. Babusiaux, M. Biermann, O.L. Creevey, D.W. Evans et al., Astron. Astrophys. {\bf 649}, A1 (2021).

\item D. Galad\'\i-Enr\'\i quez, E. Trullols, and C. Jordi, Astron. Astrophys. Suppl. Ser. {\bf 146}, 169 (2000).

\item P. Garc\'\i a-Lario, A. Manchado, S.R. Suso, S.R. Pottasch, and R. Olling, Astron. Astrophys. Suppl. Ser. {\bf 82}, 497 (1990).

\item I. Gezer, H. Van Winckel, Z. Bozkurt, Z.K. De Smedt, M. Hillen, D. Kamath, and R. Manick, MNRAS {\bf 453}, 133 (2015).

\item I. Gezer, H. Van Winckel, R. Manick, and D. Kamath, MNRAS {\bf 488}, 4033 (2019).

\item D. Gillet, A. Duquennoy, P. Bouchet, and C. Gouiffes, Astron. Astrophys. {\bf 215}, 316 (1989).

\item D. Graczyk, M. Miko\l ajewski, L. Leedj\"{a}rv, S.M. Frackowiak, J.P. Osiwa\l a, A. Puss, and 
T. Tomov, Acta Astronomica {\bf 52}, 293 (2002).

\item M.A.T. Groenewegen and M.I. Jurkovic, Astron. Astrophys. {\bf 604}, A29 (2017).

\item H.C. Harris and G. Wallerstein, Astron. J. {\bf 89}, 379 (1984).

\item N.P. Ikonnikova, M.A. Burlak, A.V. Dodin, S.Yu. Shugarov, A.A. Belinski, A.A. Fedoteva, A.M. Tatarnikov, R.J. Rudy et al., MNRAS  {\bf 530}, 1328 (2024).

\item G.H. Jacoby, D.A. Hunter, and C.A. Christian, Astrophys. J. Suppl. Ser. {\bf 56}, 257 (1984).

\item C.S. Kochanek, B.J. Shappee, K.Z. Stanek, T.W.-S. Holoien, Todd A. Thompson, J.L. Prieto, Subo Dong, J.V. Shields et al., Publ. Astron. Soc. Pacific {\bf 129}, 104502 (2017).

\item J. Lafler and T.D. Kinman, Astrophys. J. Suppl. Ser. {\bf 11}, 216 (1965).

\item A. Lebre and D. Gillet, Astron. Astrophys. {\bf 251}, 549 (1991).

\item R. Manick, H. Van Winckel, D. Kamath, S. Sekaran, and K. Kolenberg, Astron. Astrophys. {\bf 618}, A21 (2018).

\item G.-M. Oomen, H. Van Winckel, O. Pols, G. Nelemans, A. Escorza, R. Manick, D. Kamath, and C. Waelkens, Astron. Astrophys. {\bf 620}, A85 (2018).

\item G.-M. Oomen, O. Pols, H. Van Winckel, and G. Nelemans, Astron. Astrophys. {\bf 642}, A234 (2020).

\item K.R. Pollard, P.L. Cottrell, P.M. Kilmartin, and A.C. Gilmore, MNRAS {\bf 279}, 949 (1996).

\item K.R. Pollard, P.L. Cottrell, W.A. Lawson, M.D. Albrow, and W. Tobin, MNRAS {\bf 286}, 1 (1997).

\item S.A. Potanin, A.A. Belinski, A.V. Dodin, S.G. Zheltoukhov, V.Yu. Lander, K.A. Postnov, A.D. Savvin, A.M. Tatarnikov, et al., Astron. Lett. {\bf 46}, 836 (2020).

\item G.W. Preston, W. Krzeminski, J. Smak, and J.A. Williams, Astrophys. J. {\bf 137}, 401 (1963).

\item N.N. Samus’, E.V. Kazarovets, O.V. Durlevich,
N.N. Kireeva, and E.N. Pastukhova, Astron. Rep.
{\bf 61}, 80 (2017).

\item E.F. Schlafly and D.P. Finkbeiner, Astrophys. J. {\bf 737}, 103 (2011).

\item D.J. Schlegel, D.P. Finkbeiner, and M. Davis, Astrophys. J. {\bf 500}, 525 (1998).

\item B.J. Shappee, J.L. Prieto, D. Grupe, C.S. Kochanek, K.Z. Stanek, G. De Rosa, S. Mathur, Y. Zu, B.M. Peterson et al., Astrophys. J. {\bf 788}, id. 48 (2014).

\item V. Strai{\v{z}}ys, Multicolor stellar photometry, Pachart Pub. House, Tucson, 1992.

\item O. Su\'{a}rez, P. Garc\'\i a-Lario, A. Manchado, M. Manteiga, A. Ulla, and S.R. Pottasch, Astron. Astrophys. {\bf 458}, 173 (2006).

\item R. Szczerba, N. Si\'{o}dmiak, G. Stasi\'{n}ska, and J. Borkowski, Astron. Astrophys. {\bf 469}, 799 (2007).

\item G.A. Tammann, A. Sandage, and B. Reindl, Astron. Astrophys. {\bf 404}, 423 (2003).

\item V. Venkata Raman, B.G. Anandarao, P. Janardhan, and R. Pandey, MNRAS {\bf 470}, 1593 (2017).

\item S.B. Vickers, D.J. Frew, Q.A. Parker, and I.S. Boji{\v{c}}i\'{c}, MNRAS {\bf 447}, 1673 (2015).

\item G. Wallerstein, Astrophys. J. {\bf 130}, 560 (1959).

\item G. Wallerstein, Publ. Astron. Soc. Pac. {\bf 114}, 689 (2002).

\item C. Whitney, Ann. d'Astrophysique {\bf 19}, 142 (1956).

\end{enumerate}
    
\newpage

\newpage

\section{Appendix}
\setcounter{table}{0}
\renewcommand\thetable{A\arabic{table}}

\setlength{\LTcapwidth}{\dimexpr\columnwidth-1.5em\relax}
\begin{longtable}{c c c c c}
\caption{$BVR_{C}I_{C}$-photometry of IRAS~07506-0345 during 2023--2026.}
\label{tab:photometry}\\
\toprule
JD-2400000 & $B$ & $V$ & $R_C$ & $I_C$ \\
\midrule
\endfirsthead

\multicolumn{5}{c}{{\tablename\ \thetable{} -- continued}} \\
\toprule
JD-2400000 & $B$ & $V$ & $R_C$ & $I_C$ \\
\midrule
\endhead

\bottomrule
\multicolumn{5}{r}{Continued on next page} \\
\endfoot

\bottomrule
\endlastfoot

59961.396 & 16.932 & 15.877 & 15.305 & 14.830 \\
59962.326 & 16.990 & 15.959 & 15.398 & 14.936 \\
59963.443 & 17.088 & 16.041 & 15.472 & 15.015 \\
59964.322 & 17.092 & 16.113 & 15.532 & 15.070 \\
59965.305 & 17.065 & 16.126 & 15.558 & 15.122 \\
59966.351 & 16.902 & 16.055 & 15.528 & 15.126 \\
59967.305 & 16.640 & 15.908 & 15.389 & 14.988 \\
59968.350 & 16.274 & 15.625 & 15.175 & 14.775 \\
59969.367 & 16.006 & 15.411 & 14.982 & 14.615 \\
59970.250 & 15.853 & 15.297 & 14.856 & 14.501 \\
59972.398 & 15.832 & 15.225 & 14.769 & 14.394 \\
59976.273 & 15.802 & 15.060 & 14.582 & 14.168 \\
59977.335 & 15.749 & 15.015 & 14.557 & 14.144 \\
59980.355 & 15.731 & 15.023 & 14.549 & 14.113 \\
59984.319 & 15.904 & 15.160 & 14.643 & 14.186 \\
59985.032 & 16.002 & 15.177 & 14.669 & 14.240 \\
59986.153 & 16.025 & 15.235 & 14.712 & 14.282 \\
59987.052 & 16.118 & 15.264 & 14.757 & 14.315 \\
59988.031 & 16.148 & 15.335 & 14.806 & 14.369 \\
59991.287 & 16.220 & 15.523 & 15.015 & 14.580 \\
59993.277 & 16.610 & 15.586 & 15.068 & 14.644 \\
59997.249 & 16.432 & 15.644 & 15.149 & 14.871 \\
60000.349 & 16.222 & 15.531 & 15.071 & 14.708 \\
60001.309 & 16.106 & 15.462 & 15.018 & 14.625 \\
60003.229 & 16.018 & 15.286 & 14.862 & 14.558 \\
60006.298 & 15.858 & 15.225 & 14.772 & 14.410 \\
60007.364 & 15.915 & 15.151 & 14.776 & 14.440 \\
60011.334 & 15.832 & 15.118 & 14.637 & 14.255 \\
60012.222 & 15.910 & 15.125 & 14.630 & 14.208 \\
60013.272 & 15.915 & 15.119 & 14.622 & 14.197 \\
60016.250 & 15.956 & 15.100 & 14.585 & 14.201 \\
60018.197 & 15.961 & 15.088 & 14.611 & 14.349 \\
60019.265 & 16.057 & 15.159 & 14.649 & 14.219 \\
60020.233 & 16.105 & 15.206 & 14.682 & 14.245 \\
60026.273 & 16.602 & 15.587 & 15.023 & 14.554 \\
60037.199 & 16.263 & 15.550 & 15.092 & 14.712 \\
60038.224 & 16.057 & 15.435 & 14.968 & 14.623 \\
60039.203 & 15.876 & 15.315 & 14.862 & 14.480 \\
60048.251 & 15.722 & 14.936 & 14.461 & 14.086 \\
60049.210 & 15.762 & 14.968 & 14.482 & 14.069 \\
60050.281 & 15.767 & 14.981 & 14.501 & 14.130 \\
60051.208 & 15.850 & 14.999 & 14.522 & 14.111 \\
60054.219 & 16.002 & 15.132 & 14.635 & 14.204 \\
60057.219 & 16.291 & 15.347 & 14.808 & 14.387 \\
60062.228 & 16.832 & 15.814 & 15.265 & 14.842 \\
60217.573 & 15.649 & 14.873 & 14.397 & 13.971 \\
60218.583 & 15.655 & 14.898 & 14.415 & 14.005 \\
60218.586 & 15.754 & 14.894 & 14.415 & 14.005 \\
60223.567 & 16.156 & 15.127 & 14.624 & 14.190 \\
60223.569 & 15.978 & 15.133 & 14.624 & 14.190 \\
60225.556 & 16.066 & 15.226 & 14.727 & 14.290 \\
60225.558 & 16.082 & 15.225 & 14.727 & 14.290 \\
60229.571 & 16.340 & 15.486 & 14.950 & 14.519 \\
60230.525 & 16.422 & 15.561 & 15.022 & 14.595 \\
60230.527 & 16.418 & 15.561 & 15.022 & 14.595 \\
60232.538 & 16.518 & 15.671 & 15.153 & 14.712 \\
60232.540 & 16.524 & 15.670 & 15.153 & 14.712 \\
60239.534 & 15.900 & 15.344 & 14.924 & 14.572 \\
60239.536 & 15.910 & 15.344 & 14.924 & 14.572 \\
60240.544 & 15.780 & 15.233 & 14.820 & 14.478 \\
60240.546 & 15.784 & 15.233 & 14.820 & 14.478 \\
60245.526 & 15.923 & 15.148 & 14.689 & 14.327 \\
60246.515 & 15.873 & 15.147 & 14.668 & 14.269 \\
60246.517 & 15.898 & 15.146 & 14.668 & 14.269 \\
60248.560 & 15.880 & 15.084 & 14.599 & 14.183 \\
60248.562 & 15.859 & 15.085 & 14.599 & 14.183 \\
60249.551 & 15.866 & 15.063 & 14.582 & 14.167 \\
60249.553 & 15.856 & 15.063 & 14.582 & 14.167 \\
60253.511 & 15.974 & 15.083 & 14.570 & 14.168 \\
60254.541 & 15.997 & 15.107 & 14.609 & 14.219 \\
60255.562 & 16.044 & 15.153 & 14.632 & 14.252 \\
60255.564 & 16.051 & 15.153 & 14.632 & 14.252 \\
60256.539 & 16.112 & 15.192 & 14.674 & 14.277 \\
60256.541 & 16.112 & 15.192 & 14.674 & 14.277 \\
60259.548 & 16.353 & 15.369 & 14.819 & 14.374 \\
60260.487 & 16.438 & 15.428 & 14.871 & 14.450 \\
60263.524 & 16.738 & 15.617 & 15.067 & 14.671 \\
60277.480 & 15.735 & 15.155 & 14.677 & 14.347 \\
60277.481 & 15.753 & 15.154 & 14.677 & 14.347 \\
60281.482 & 15.779 & 15.064 & 14.582 & 14.187 \\
60282.424 & 15.742 & 15.026 & 14.554 & 14.165 \\
60282.426 & 15.778 & 15.025 & 14.554 & 14.165 \\
60283.467 & 15.736 & 14.989 & 14.521 & 14.131 \\
60283.469 & 15.734 & 14.989 & 14.521 & 14.131 \\
60287.515 & 15.764 & 14.957 & 14.476 & 14.098 \\
60288.434 & 15.816 & 14.981 & 14.482 & 14.261 \\
60288.512 & 15.805 & 15.002 & 14.505 & 14.102 \\
60293.434 & 16.128 & 15.204 & 14.704 & 14.266 \\
60294.481 & 16.216 & 15.275 & 14.744 & 14.303 \\
60296.380 & 16.331 & 15.374 & 14.846 & 14.423 \\
60298.382 & 16.531 & 15.546 & 15.003 & 14.553 \\
60299.394 & 16.619 & 15.629 & 15.088 & 14.634 \\
60303.366 & 16.778 & 15.915 & 15.363 & 15.117 \\
60305.418 & 16.716 & 15.848 & 15.362 & 14.929 \\
60305.420 & 16.887 & 15.842 & 15.362 & 14.929 \\
60308.358 & 16.129 & 15.472 & 15.000 & 14.608 \\
60310.366 & 15.730 & 15.155 & 14.759 & 14.423 \\
60311.369 & 15.686 & 15.120 & 14.710 & 14.374 \\
60313.434 & 15.743 & 15.079 & 14.627 & 14.217 \\
60315.350 & 15.696 & 14.991 & 14.516 & 14.146 \\
60316.356 & 15.692 & 14.964 & 14.476 & 14.078 \\
60320.328 & 15.770 & 14.957 & 14.467 & 14.079 \\
60321.386 & 15.804 & 14.979 & 14.487 & 14.086 \\
60324.332 & 15.965 & 15.081 & 14.583 & 14.164 \\
60326.512 & 16.115 & 15.226 & 14.693 & 14.298 \\
60327.354 & 16.198 & 15.267 & 14.737 & 14.305 \\
60328.315 & 16.262 & 15.347 & 14.792 & 14.388 \\
60328.377 & 16.292 & 15.343 & 14.795 & 14.444 \\
60330.286 & 16.518 & 15.471 & 14.952 & 14.564 \\
60334.375 & 16.895 & 15.885 & 15.267 & 14.846 \\
60340.410 & 16.498 & 15.746 & 15.280 & 14.886 \\
60342.362 & 15.906 & 15.349 & 14.935 & 14.577 \\
60345.217 & 15.548 & 15.028 & 14.633 & 14.312 \\
60348.241 & 15.690 & 15.025 & 14.580 & 14.195 \\
60350.243 & 15.716 & 14.983 & 14.519 & 14.151 \\
60351.259 & 15.726 & 14.983 & 14.511 & 14.092 \\
60353.230 & 15.759 & 14.979 & 14.498 & 14.091 \\
60355.261 & 15.882 & 15.005 & 14.556 & 14.197 \\
60364.267 & 16.575 & 15.562 & 14.994 & 14.536 \\
60365.292 & 16.730 & 15.648 & 15.084 & 14.622 \\
60366.356 & 16.819 & 15.724 & 15.164 & 14.775 \\
60367.277 & 16.839 & 15.765 & 15.242 & 14.786 \\
60368.341 & 16.896 & 15.876 & 15.316 & 14.830 \\
60369.311 & 16.968 & 15.935 & 15.361 & 14.919 \\
60370.276 & 16.950 & 15.974 & 15.404 & 14.952 \\
60371.226 & 16.896 & 15.972 & 15.417 & 14.920 \\
60373.233 & 16.529 & 15.751 & 15.240 & 14.899 \\
60374.285 & 16.238 & 15.628 & 15.076 & 14.739 \\
60375.229 & 16.065 & 15.401 & 14.959 & 14.579 \\
60376.326 & 15.837 & 15.260 & 14.830 & 14.466 \\
60380.366 & 15.789 & 15.108 & 14.649 & 14.292 \\
60382.241 & 15.790 & 15.062 & 14.585 & 14.179 \\
60384.318 & 15.766 & 14.983 & 14.515 & 14.121 \\
60391.301 & 15.829 & 15.067 & 14.558 & 14.132 \\
60396.218 & 16.072 & 15.242 & 14.749 & 14.347 \\
60398.199 & 16.202 & 15.354 & 14.842 & 14.409 \\
60400.253 & 16.248 & 15.515 & 15.014 & 14.726 \\
60401.203 & 16.348 & 15.522 & 15.008 & 14.592 \\
60402.209 & 16.367 & 15.574 & 15.064 & 14.663 \\
60403.204 & 16.446 & 15.617 & 15.115 & 14.703 \\
60404.227 & 16.473 & 15.660 & 15.161 & 14.837 \\
60405.208 & 16.476 & 15.672 & 15.188 & 14.773 \\
60410.260 & 15.937 & 15.330 & 14.916 & 14.551 \\
60413.211 & 15.912 & 15.263 & 14.791 & 14.418 \\
60421.234 & 15.912 & 14.998 & 14.519 & 14.127 \\
60424.222 & 16.014 & 15.049 & 14.552 & 14.144 \\
60426.226 & 15.845 & 15.204 & 14.672 & 14.455 \\
60429.227 & 16.260 & 15.297 & 14.754 & 14.322 \\
60595.584 & 16.064 & 15.201 & 14.685 & 14.245 \\
60606.602 & 17.085 & 16.113 & 15.554 & 15.116 \\
60607.561 & 17.051 & 16.129 & 15.575 & 15.128 \\
60610.601 & 16.231 & 15.586 & 15.153 & 14.776 \\
60616.547 & 15.725 & 15.069 & 14.620 & 14.246 \\
60622.587 & 15.818 & 15.017 & 14.523 & 14.113 \\
60624.580 & 15.894 & 15.063 & 14.562 & 14.134 \\
60625.562 & 15.940 & 15.102 & 14.596 & 14.169 \\
60626.591 & 16.003 & 15.132 & 14.629 & 14.204 \\
60628.559 & 16.160 & 15.250 & 14.729 & 14.295 \\
60631.561 & 16.488 & 15.482 & 14.929 & 14.459 \\
60634.530 & 16.841 & 15.759 & 15.195 & 14.733 \\
60635.561 & 16.935 & 15.861 & 15.292 & 14.802 \\
60637.518 & 17.064 & 16.029 & 15.450 & 14.998 \\
60638.536 & 17.089 & 16.075 & 15.523 & 15.089 \\
60645.533 & 15.589 & 15.074 & 14.691 & 14.361 \\
60646.492 & 15.593 & 15.043 & 14.650 & 14.307 \\
60647.553 & 15.646 & 15.046 & 14.634 & 14.262 \\
60648.515 & 15.702 & 15.060 & 14.623 & 14.229 \\
60649.508 & 15.728 & 15.051 & 14.605 & 14.207 \\
60651.492 & 15.774 & 15.042 & 14.575 & 14.162 \\
60652.484 & 15.803 & 15.051 & 14.571 & 14.170 \\
60653.457 & 15.830 & 15.040 & 14.565 & 14.162 \\
60654.564 & 15.846 & 15.059 & 14.570 & 14.162 \\
60655.483 & 15.871 & 15.072 & 14.587 & 14.168 \\
60659.460 & 16.136 & 15.227 & 14.705 & 14.279 \\
60659.483 & 16.114 & 15.236 & 14.706 & 14.276 \\
60664.531 & 16.462 & 15.511 & 14.954 & 14.487 \\
60665.416 & 16.513 & 15.554 & 15.006 & 14.544 \\
60666.482 & 16.591 & 15.626 & 15.070 & 14.628 \\
60669.440 & 16.781 & 15.808 & 15.240 & 14.749 \\
60671.427 & 16.769 & 15.777 & 15.311 & 14.962 \\
60675.425 & 16.342 & 15.618 & 15.146 & 14.763 \\
60676.356 & 16.128 & 15.487 & 15.038 & 14.680 \\
60677.411 & 15.937 & 15.325 & 14.895 & 14.548 \\
60678.425 & 15.806 & 15.214 & 14.803 & 14.442 \\
60679.423 & 15.745 & 15.158 & 14.743 & 14.390 \\
60680.421 & 15.777 & 15.154 & 14.732 & 14.391 \\
60682.473 & 15.842 & 15.162 & 14.710 & 14.328 \\
60684.490 & 15.900 & 15.148 & 14.673 & 14.264 \\
60685.377 & 15.897 & 15.124 & 14.646 & 14.228 \\
60686.376 & 15.884 & 15.097 & 14.618 & 14.194 \\
60691.382 & 15.940 & 15.110 & 14.616 & 14.199 \\
60694.390 & 16.075 & 15.203 & 14.693 & 14.295 \\
60696.405 & 16.176 & 15.292 & 14.761 & 14.359 \\
60697.398 & 16.136 & 15.340 & 14.806 & 14.357 \\
60698.307 & 16.285 & 15.385 & 14.863 & 14.456 \\
60700.378 & 16.437 & 15.504 & 14.971 & 14.516 \\
60702.401 & 16.565 & 15.621 & 15.091 & 14.614 \\
60703.362 & 16.601 & 15.678 & 15.140 & 14.688 \\
60704.363 & 16.668 & 15.738 & 15.206 & 14.764 \\
60706.334 & 16.723 & 15.832 & 15.303 & 14.866 \\
60707.319 & 16.710 & 15.844 & 15.326 & 14.882 \\
60708.350 & 16.665 & 15.819 & 15.327 & 14.935 \\
60709.308 & 16.556 & 15.784 & 15.283 & 14.867 \\
60715.261 & 15.916 & 15.255 & 14.813 & 14.442 \\
60716.350 & 15.946 & 15.223 & 14.777 & 14.383 \\
60717.299 & 15.935 & 15.181 & 14.708 & 14.296 \\
60718.342 & 15.917 & 15.137 & 14.677 & 14.267 \\
60719.308 & 15.900 & 15.120 & 14.614 & 14.206 \\
60720.303 & 15.959 & 15.010 & 14.575 & 14.162 \\
60722.417 & 15.828 & 15.038 & 14.533 & 14.122 \\
60729.262 & 16.139 & 15.223 & 14.715 & 14.279 \\
60735.321 & 16.733 & 15.696 & 15.121 & 14.666 \\
60736.263 & 16.822 & 15.781 & 15.204 & 14.736 \\
60737.228 & 16.910 & 15.871 & 15.283 & 14.783 \\
60739.256 & 17.109 & 16.009 & 15.449 & 14.994 \\
60742.273 & 16.981 & 16.063 & 15.502 & 15.050 \\
60743.259 & 16.768 & 15.943 & 15.432 & 15.002 \\
60744.288 & 16.494 & 15.749 & 15.280 & 14.879 \\
60745.239 & 16.159 & 15.530 & 15.075 & 14.721 \\
60746.231 & 15.866 & 15.291 & 14.889 & 14.538 \\
60747.318 & 15.725 & 15.172 & 14.767 & 14.420 \\
60748.261 & 15.711 & 15.155 & 14.725 & 14.379 \\
60751.246 & 15.768 & 15.089 & 14.626 & 14.204 \\
60759.210 & 15.890 & 15.084 & 14.597 & 14.153 \\
60760.279 & 15.961 & 15.137 & 14.633 & 14.204 \\
60763.225 & 16.175 & 15.279 & 14.771 & 14.340 \\
60764.250 & 16.239 & 15.342 & 14.825 & 14.383 \\
60784.240 & 15.820 & 15.133 & 14.674 & 14.326 \\
60785.234 & 15.819 & 15.101 & 14.641 & 14.245 \\
60967.594 & 16.417 & 15.454 & 14.915 & 14.464 \\
60975.586 & 17.131 & 16.115 & 15.549 & 15.062 \\
60979.570 & 16.510 & 15.774 & 15.288 & 14.882 \\
60980.580 & 16.244 & 15.575 & 15.128 & 14.746 \\
60982.577 & 15.834 & 15.262 & 14.838 & 14.481 \\
60985.581 & 15.836 & 15.176 & 14.718 & 14.340 \\
60986.536 & 15.838 & 15.146 & 14.671 & 14.281 \\
60987.544 & 15.830 & 15.117 & 14.629 & 14.214 \\
60996.535 & 15.990 & 15.163 & 14.659 & 14.218 \\
60997.500 & 16.044 & 15.198 & 14.696 & 14.253 \\
60998.502 & 16.098 & 15.247 & 14.742 & 14.299 \\
61001.547 & 16.277 & 15.387 & 14.864 & 14.414 \\
61002.443 & 16.318 & 15.431 & 14.905 & 14.468 \\
61007.505 & 16.544 & 15.680 & 15.166 & 14.715 \\
61009.489 & 16.522 & 15.724 & 15.219 & 14.793 \\
61013.518 & 16.177 & 15.543 & 15.094 & 14.707 \\
61014.509 & 16.058 & 15.439 & 15.010 & 14.613 \\
61015.516 & 15.965 & 15.367 & 14.945 & 14.559 \\
61026.457 & 16.016 & 15.175 & 14.665 & 14.215 \\
61028.456 & 16.061 & 15.172 & 14.667 & 14.220 \\
61029.466 & 16.084 & 15.197 & 14.681 & 14.239 \\
61030.414 & 16.120 & 15.221 & 14.705 & 14.258 \\
61035.475 & 16.489 & 15.474 & 14.910 & 14.438 \\
61038.437 & 16.719 & 15.658 & 15.082 & 14.604 \\
61043.408 & 17.058 & 16.014 & 15.428 & 14.946 \\
61045.391 & 17.069 & 16.051 & 15.491 & 15.020 \\
61048.430 & 16.675 & 15.863 & 15.346 & 14.917 \\
61053.346 & 15.845 & 15.216 & 14.775 & 14.422 \\
61055.395 & 15.802 & 15.132 & 14.675 & 14.276 \\
61056.386 & 15.741 & 15.059 & 14.605 & 14.204 \\
61057.442 & 15.692 & 14.999 & 14.537 & 14.129 \\
61060.448 & 15.684 & 14.937 & 14.475 & 14.052 \\
61061.310 & 15.688 & 14.952 & 14.460 & 14.045 \\
61062.380 & 15.760 & 14.981 & 14.503 & 14.079 \\
61066.402 & 15.982 & 15.154 & 14.653 & 14.206 \\
61068.356 & 16.094 & 15.234 & 14.732 & 14.281 \\
61069.383 & 16.169 & 15.300 & 14.778 & 14.327 \\
61073.433 & 16.576 & 15.571 & 15.046 & 14.579 \\
61079.313 & 16.889 & 15.993 & 15.468 & 15.026 \\
61082.322 & 16.537 & 15.839 & 15.376 & 14.964 \\
61086.397 & 15.798 & 15.238 & 14.824 & 14.462 \\
61087.235 & 15.810 & 15.224 & 14.790 & 14.407 \\
61089.389 & 15.850 & 15.157 & 14.695 & 14.296 \\
61091.289 & 15.820 & 15.077 & 14.604 & 14.186 \\
61092.282 & 15.824 & 15.055 & 14.567 & 14.145 \\
61097.378 & 15.972 & 15.108 & 14.611 & 14.182 \\
61102.294 & 16.310 & 15.355 & 14.835 & 14.370 \\
61109.360 & 17.169 & 16.043 & 15.450 & 14.960 \\
61110.202 & 17.210 & 16.117 & 15.509 & 15.043 \\
61112.272 & 17.273 & 16.217 & 15.611 & 15.134 \\
61113.270 & 17.171 & 16.172 & 15.605 & 15.137 \\
61114.241 & 16.969 & 16.055 & 15.522 & 15.056 \\
\end{longtable}

\end{document}